\documentstyle[11pt,newpasp,twoside,epsf]{article}
\def\arg#1{{\it#1\/}}

\def\edcomment#1{\iffalse\marginpar{\raggedright\sl#1\/}\else\relax\fi}
\reversemarginpar

\begin{document}
\title{Determining the Metallicity of Cepheid Stars in the SMC, LMC and the
Galaxy} \author{Marta Mottini}
 \affil{European Southern Observatory, K. Schwarzschild strasse 2,
 D-85748 Garching b. M{\"u}nchen, Germany} \author{Francesca Primas}
 \affil{European Southern Observatory, K. Schwarzschild strasse 2,
 D-85748 Garching b. M{\"u}nchen, Germany} \author{Martino Romaniello}
 \affil{European Southern Observatory, K. Schwarzschild strasse 2,
 D-85748 Garching b. M{\"u}nchen, Germany} \author{Martin Groenewegen}
 \affil{Instituut voor Sterrenkunde, Celestijnenlaan 200B, B-3001
 Leuven, Belgium}

\begin{abstract}

The Cepheid Period-Luminosity relation is unquestionably one of the most
powerful tools at our disposal for determining the extragalactic distance
scale. While significant progress has been made in the past few years
towards its understanding and characterisation, both on the observational
(e.g. the HST Key Project) and theoretical (e.g. non-linear pulsation models,
non-LTE atmospheres etc.) sides, the debate on the influence that chemical
composition may have on the Period-Luminosity relation is still unsettled.
Current estimates lead to differences in the distance as large as 15\%,
effectively limiting the accuracy of Cepheids as distance indicators.

To further tackle this problem, we have obtained high resolution spectra of
a large sample of Cepheids in our Galaxy and the Magellanic Clouds. The
superb quality of the data allow us to probe the detailed effects of
chemical composition (alpha, iron-group, and heavy elements)  over more than
a factor of ten in metallicity. Here, we present the first preliminary
results of the analysis of iron abundances in a sub-sample of Cepheids.

\end{abstract}

\section{Introduction}

The question if (and by how much) metallicity has any effect on the Cepheid
Period-Luminosity (PL) relation is far from being settled. Recent theoretical
studies give different results: some authors claim that metallicity has
negligible effects on the PL relation (Baraffe \& Alibert 2001, Alibert et
al. 1999; Sandage et al. 1999); others find that there is a significant
dependence of the PL relation on metallicity (Bono et al. 1999; Caputo et
al. 2000; Fiorentino et al. 2002).

Although most of the observational efforts aiming at deriving the chemical
composition of Cepheids and its effects on the PL relation have used indirect
means, like the measurement of O~II in HII regions (e.g. Beaulieu et al. 1997;
Sasselov et al. 1997; Kennicutt et al. 1998; Macri et al. 2001), some of them
have focused on a full and direct chemical analysis (Fry \& Carney 1997;
Andrievsky et al. 2002a,b; Luck et al. 2003). The goal of our project is to
follow this second approach, i.e. to make a direct determination of the
metallicity in a newly observed sample of Cepheid stars.

\section{The Data Sample}

The data-set has been assembled according to the following selection
criteria: a large coverage of metallicities (hence, we selected stars from
three different galaxies: the SMC, the LMC, and our own); the availability
of well-determined intensity-mean magnitudes in the B, V, J, H, and K bands
(cf Berdnikov \& Caldwell 2001; Groenewegen 1999; Laney \& Stobie 1994, Laney
\& Stobie 1992); a wide range of periods in order to cover the entire PL
relation (hence we chose objects from 2 to 98 days of period); the
availability of Hipparcos parallaxes for all Galactic Cepheids.

Our sample includes a total of 76 stars: 40 Galactic, 22 LMC, and 14 SMC
Cepheids. The Magellanic Cepheids have been observed with UVES (R=40,000,
spectral coverage: 480--680~nm) at the ESO-VLT and have typical S/N of the
order of 50 on the EEV CCD (i.e. for 480$\leq \lambda \leq$580~nm) and
$\simeq$70 on the MIT CCD (i.e. for 580$\leq \lambda \leq$ 680~nm). The
spectra of the Galactic sample, instead, have been collected with FEROS
(R=45,000, spectral coverage: 380--880~nm) at the ESO 1.5m telescope in La
Silla. The typical S/N of these spectra is $\simeq$70 up to $\lambda$=500~nm, 
but larger than 150 in the red.

\section{The Analysis}

After the spectra were reduced with the help of the FEROS and UVES Data
Reduction Software pipelines, our main goal has been to develop a reliable
analytical procedure in order to accurately determine the iron abundance
of these stars.

First, we carefully assembled a reliable list of iron lines (both neutral and
ionized). We started from a preliminary selection from the VALD database
(Kupka et al. 1999; Piskunov et al. 1995) choosing lines between 450 and
650~nm, for effective temperatures typical of Cepheids (4500-6500~K). Every
line profile was then visually inspected on the observed spectra in order to
detect (and eliminate) those lines affected by other elemental blends. 

Secondly we measured the equivalent widths (EWs) of all the lines assembled
as described above. To do this, we tested two methods: we used the gaussian
fitting option offered by the task {\em splot} in IRAF; we then developed an
IDL routine (based on the {\em line\_width} widget available in the FUSE IDL
suite of tools) in order to speed up the process, but keeping the whole
procedure interactive.

\begin{figure}
\plotfiddle{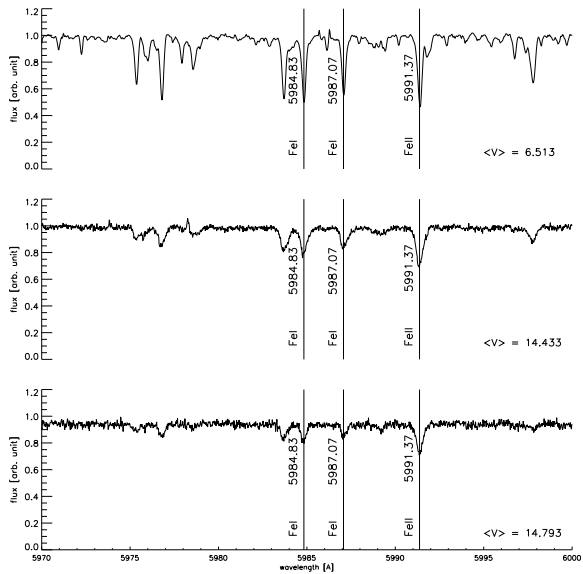}{7cm}{0}{38}{38}{-140}{-40}
\caption{The same spectral range is shown for three different Cepheids of our sample, characterized by similar periods (top: Milky Way, middle: LMC, bottom: SMC).}
\end{figure}

With a final list of reliable equivalent widths in hand, we then made use of
the code WIDTH9 (Kurucz 1993) for the computation of our final iron
abundances. For this purpose, we chose the Kurucz grid of LTE model
atmospheres (Kurucz 1993). However, before being able to derive
[Fe/H]\footnote{where [Fe/H]=log(Fe/H)$_*$ - log(Fe/H)$_{\odot}$} values
from our measured EWs, one clearly needs to know the stellar parameters of
the stars, i.e. their effective temperature, gravity, and microtubulent
velocity. We derived a first guess set of stellar parameters from the
photometry available in the literature (see references given in Section 2)
and determine spectroscopically the final set to be used in the
analysis. This is an interactive process that requires running the code
WIDTH9, with our final list of EWs and different model atmospheres, in order
to find which model is able to minimize simultaneously the slope of
log$\epsilon$(Fe) {\it vs} EW and {\it vs} the excitation potential (these
constrain the microturbulence and the temperature respectively) while keeping
the ionization balance (i.e. the abundance derived from Fe~I and from Fe~II
must agree)to constrain the gravity. Once such model was identified for each individual star,
we were finally able to compute the iron abundance of all our stars.

\section{Preliminary results and Future Developments}

We have determined the iron abundances for a sub-sample of stars: 14 Galactic,
 14 LMC and 22 SMC Cepheids. Figure 1 of Romaniello et al. (this volume)
summarises this preliminary set of results. 

Despite the confidence we have gained in the analytical steps tested sofar,
we still have a lot of work in front of us. In the very near future, we want to
optimize our line-list (both for the number of lines to be used and for a
careful check of their gf-values), to complete the analysis for the entire
sample, and to determine iron abundances also via spectrum synthesis. A little
bit more into the future, we plan to perform the full chemical analysis of the
entire data-set (with special attention devoted to the $\alpha$ and to the
heavy elements).

\end{document}